\documentclass[11pt,]{article}
\usepackage{lmodern}
\usepackage{amssymb,amsmath}
\usepackage{ifxetex,ifluatex}
\usepackage{fixltx2e} 
\ifnum 0\ifxetex 1\fi\ifluatex 1\fi=0 
  \usepackage[T1]{fontenc}
  \usepackage[utf8]{inputenc}
\else 
  \ifxetex
    \usepackage{mathspec}
  \else
    \usepackage{fontspec}
  \fi
  \defaultfontfeatures{Ligatures=TeX,Scale=MatchLowercase}
\fi
\IfFileExists{upquote.sty}{\usepackage{upquote}}{}
\IfFileExists{microtype.sty}{%
\usepackage{microtype}
\UseMicrotypeSet[protrusion]{basicmath} 
}{}
\usepackage[margin=1in]{geometry}
\usepackage{hyperref}
\hypersetup{unicode=true,
            pdfborder={0 0 0},
            breaklinks=true}
\usepackage{color}
\usepackage{fancyvrb}

\DefineVerbatimEnvironment{Highlighting}{Verbatim}{commandchars=\\\{\}}
\usepackage{framed}
\definecolor{shadecolor}{RGB}{248,248,248}
\newenvironment{Shaded}{\begin{snugshade}}{\end{snugshade}}
\newcommand{\KeywordTok}[1]{\textcolor[rgb]{0.13,0.29,0.53}{\textbf{#1}}}
\newcommand{\DataTypeTok}[1]{\textcolor[rgb]{0.13,0.29,0.53}{#1}}
\newcommand{\DecValTok}[1]{\textcolor[rgb]{0.00,0.00,0.81}{#1}}

\newcommand{\StringTok}[1]{\textcolor[rgb]{0.31,0.60,0.02}{#1}}

\newcommand{\OtherTok}[1]{\textcolor[rgb]{0.56,0.35,0.01}{#1}}

\newcommand{\OperatorTok}[1]{\textcolor[rgb]{0.81,0.36,0.00}{\textbf{#1}}}

\newcommand{\NormalTok}[1]{#1}
\usepackage{longtable,booktabs}
\usepackage{graphicx,grffile}
\makeatletter
\def\maxwidth{\ifdim\Gin@nat@width>\linewidth\linewidth\else\Gin@nat@width\fi}
\def\maxheight{\ifdim\Gin@nat@height>\textheight\textheight\else\Gin@nat@height\fi}
\makeatother
\setkeys{Gin}{width=\maxwidth,height=\maxheight,keepaspectratio}
\IfFileExists{parskip.sty}{%
\usepackage{parskip}
}{
\setlength{\parindent}{0pt}
\setlength{\parskip}{6pt plus 2pt minus 1pt}
}
\ifx\paragraph\undefined\else
\let\oldparagraph\paragraph
\renewcommand{\paragraph}[1]{\oldparagraph{#1}\mbox{}}
\fi
\ifx\subparagraph\undefined\else
\let\oldsubparagraph\subparagraph
\renewcommand{\subparagraph}[1]{\oldsubparagraph{#1}\mbox{}}
\fi

\let\rmarkdownfootnote\footnote%
\def\footnote{\protect\rmarkdownfootnote}

\usepackage{titling}

  \title{A novel high-power test for continuous outcomes truncated by death}
  \pretitle{\vspace{\droptitle}\centering\huge}
  \posttitle{\par}
  \author{Andreas Kryger Jensen\textsuperscript{1} and Theis
Lange\textsuperscript{1,2}\\
1. Section of Biostatistics, Dep. of Public Health, University of
Copenhagen\\
2. Center for Statistical Science, Peking University, Beijing, China}
  \preauthor{\centering\large\emph}
  \postauthor{\par}
  \predate{\centering\large\emph}
  \postdate{\par}
  \date{27 October, 2019}

\DeclareMathOperator*{\E}{E}
\DeclareMathOperator*{\logit}{logit}
\usepackage[labelfont=bf]{caption}
\usepackage{amsthm}
\newtheorem{proposition}{Proposition}

\begin{document}
\maketitle

\begin{abstract}
Patient reported outcomes including quality of life (QoL) assessments are increasingly being included as either primary or secondary outcomes in randomized controlled trials. While making the outcomes more relevant for patients it entails a challenge in cases where death or a similar event makes the outcome of interest undefined. A pragmatic - and much used - solution is to assign diseased patient with the lowest possible QoL score. This makes medical sense, but creates a statistical problem since traditional tests such as t-tests or Wilcox tests potentially looses large amounts of statistical power. In this paper we propose a novel test that can keep the medical relevant composite outcome, but preserve full statistical power. The test is also applicable in other situations where a specific value (say 0 days alive outside hospitals) encodes a special meaning. The test is implemented in an R package which is available for download. 
\end{abstract}

\begin{center}
\end{center}

\section{Introduction}\label{introduction}

In medical intervention research including in particular randomized
controlled trials (RCTs) there is a trend towards increased use of
patient reported outcomes; Quality of Life (QoL) scores are one of the
most prominent of these. Established statistical practice is to compare
these between treatment groups using non-parametric tests such as Wilcox
since distributions are rarely normal. When all patients are alive and
able and willing to provide QoL scores at the scheduled measurement time
this procedure works very well. Not all clinical settings, however, will
have all participants alive at time of scheduled assessment of QoL. This
is particular true in trials within intensive care where mortality
approximating 30\% are not uncommon and which is the setting that have
motivated this work. A pragmatic solution which is gaining popularity is
to simply give the diseased patient the lowest possible score and then
use a Wilcoxon test or similar to compare groups. This paper will
demonstrate that this approach can lead to dramatic loss of statistical
power. The paper will also presents an alternative and novel statistical
test which avoids the power loss. The novel test procedure is
implemented in R and made publicly available.

The key-insight in the proposed test is to incorporate that the outcome
is actually a two-dimensional outcome of a very special type and that
the constructed combined outcome follows a continuous-singular mixture
distribution. This unusual distribution is why one cannot resort to
non-parametric Wilcoxon (Mann-Whitney) tests since the singular
component of the distribution of the combined outcome will get reduced
to simple ties. It is noted that the handling of ties in standard
statistical software varies and is opaque. However, the handling of ties
is not the main reason why Wilcox suffers power loss. The main reason is
that the null-hypothesis in these Wilcoxon type tests (stochastic
domination) does not handle the empirical fact that treatments might
influence mortality and QoL differently.

As an alternative we propose to model the binary component (i.e.,
survival) and the continuous part (i.e., actual QoL) separately but to
conduct a single test for no treatment effect on either. We can thus
provide a single p-value for the hypothesis of no treatment effect on
the extended QoL where death is given the lowest possible score. To
accommodate potential non-normality of the recorded QoL scores we
include both parametric and a semi-parametric tests where the latter is
as widely applicable as the Wilcox test. Simulations indicate that the
semi-parametric is preferred as is greatly extends applicability of the
test procedure at a very low price in terms of reduced power. Both test
procedures will provide effect estimates of mean differences between
treatment groups based on the combined outcome along with corresponding
confidence intervals. This is also an added benefit compared to the
Wilcox-type based testing approach.

It should be noted that while we in this paper exemplifies the procedure
using QoL tests and mortality the method is applicable in any setting
where a single value of a combined outcome has a special interpretation
compared to an otherwise continuous scale. As an example, again from
intensive care research, consider the outcome ``days alive and out of
hospital within 90 days from randomization''. Here in-hospital
fatalities will all have the value 0, while everybody else will have
outcomes ranging from 0 to 90. Such outcomes are also routinely analyzed
using Wilcox-type test. Our proposed test would increase power while
also providing a mean effect estimate along with confidence bands. Note
also that the developed mathematical method allows straightforwardly for
the inclusion of confounders variables. The method itself is therefore
just as applicable in non-randomized studies or epidemiological studies
in general.

The rest of the paper is structured as follows. The next section
introduces the mathematical setup as well as our novel test procedure.
Section 3 describes the R implementation and Section 4 presents a
simulation study illustrating the substantial power gains. Finally,
Section 5 discusses. Mathematical proofs are contained in the appendix.

\section{Method}\label{sec:method}

We consider random variables \((Y, A, R, X)\) where \(Y\) is the
continuous outcome variable, \(A\) is a binary variable being equal to
\(1\) if \(Y\) is observed and equal to \(0\) if \(Y\) is
unobserved/undefined, \(R\) is a binary treatment indicator, and \(X\)
is a \(p\)-dimensional vector of baseline covariates. The statistical
objective is to describe the distribution of
\(((Y \mid A = 1), A) \mid R, X\) where the primary hypothesis of no
treatment effect is given by \(((Y \mid A = 1), A) \perp R \mid X\).

Even though the outcome is bi-variate, a combined outcome is often used
in practice. The combined outcome is derived such that it is equal to
\(Y\) if \(Y\) is observed, and otherwise it is equal to some
predetermined value. We can write this combined outcome as\\

\begin{align}
  \widetilde{Y} =  Y 1_{A = 1} + \mathcal{E} 1_{A = 0}\label{eq:combinedOutcome}
\end{align}

where \(\mathcal{E}\) is a fixed atom assigned as the outcome value when
\(Y\) is unobserved. The semi-continuous distribution of
\(\widetilde{Y}\) is therefore a probabilistic mixture of a singular
distribution at \(\mathcal{E}\) and a continuous distribution over the
domain of the random variable \(Y \mid A = 1\). Thus the statistical
challenge be be rephrased as assessing if the treatment (\(R\)) affects
the \emph{distribution} of \(\widetilde{Y}\).

The conditional expected value of the combined outcome in Equation
(\ref{eq:combinedOutcome}) is given by

\begin{align*}
  E[\widetilde{Y} \mid R, X] = E[Y \mid R, X, A=1]P(A = 1 \mid R, X) + \mathcal{E}P(A = 0 \mid R, X)
\end{align*}

From this expression one can show that a treatment comparison expressed
in terms of a contrast of the expectation of \(\widetilde{Y}\) can be
zero even though the distribution of \(((Y \mid A=1), A)\) depends on
\(R\). To see this, let \(E[Y \mid A=1, R = 0, X] = \mu(X)\) and
\(E[Y \mid A=1, R = 1, X] = \mu(X) + \mu_{\delta}(X)\) and assume
without loss of generalization that \(\mathcal{E} = 0\). Then the
average treatment effect for the combined outcome conditional on
baseline covariates is

\begin{align}
  \Delta(X) &= E[\widetilde{Y} \mid R = 1, X] - E[\widetilde{Y} \mid R = 0, X]\label{eq:Delta}\\
        &= \mu(X)\left[P(A = 1 \mid R=1,X) - P(A = 1 \mid R = 0, X)\right] + \mu_{\delta}(X) P(A = 1 \mid R = 1, X)\nonumber
\end{align}

If we auspiciously let
\(\mu_{\delta}(X) = \mu(X)\left[P(A = 1 \mid R = 0, X) P(A = 1 \mid R = 1, X)^{-1} - 1\right]\)
then \(\Delta(X) = 0\) for all values of \(X\) even though
\(\mu_{\delta}(X) \ne 0\) and
\(P(A = 1 \mid R = 0, X) \ne P(A = 1 \mid R = 1, X)\). On the other
hand, if \(\mu_\delta(X) = 0\) and
\(P(A = 1 \mid R=1,X) = P(A = 1 \mid R = 0, X)\) then \(\Delta(X)\) is
necessarily equal to zero. This illustrates that a significance test of
no treatment effect must have two degrees of freedom. Such a test which
we develop in the subsequent sections is conceptually difference than a
test for the null-hypothesis \(\Delta(X) = 0\). Noe that in RCTs one
would typically not include any variables \(X\) since these are balanced
by design.

\subsection{Likelihood ratio test}\label{likelihood-ratio-test}

We propose to test the null-hypothesis of no treatment effect on the
combined outcome by a likelihood ratio test of the joint distribution of
\(Y_i \mid A_i\) and \(A_i\). This has the advantage that it increases
the efficiency compared the Wilcoxon test and it yields a single p-value
appropriate for testing a primary outcome in a clinical trial. Let
\((Y_i, A_i, R_i, X_i)\) for \(i = 1, \ldots, n\) be independent and
identically distributed random variables. We can write our model in a
general form as

\begin{align*}
  Y_i \mid A_i = 1, R_i, X_i &\sim F(Y_i; \mu_i(R_i, X_i), \Psi)\\
  A_i \mid R_i, X_i &\sim \text{Bernoulli}(A_i; \pi_i(R_i, X_i))  
\end{align*}

for some mean functions \(\pi_i\) and \(\mu_i\) and a distribution
function \(F\) characterizing the distribution of the observed
continuous outcomes with possible nuisance parameter \(\Psi\). The joint
likelihood function for the combined outcome conditional on treatment
and baseline covariates is

\begin{align*}
 L_n(\mu_i(R_i, X_i), \pi_i(R_i, X_i), \Psi) &= \prod_{i=1}^n f(Y_i; \mu_i(R_i, X_i), \Psi)^{A_i}P(A_i = a_i; \pi_i(R_i, X_i))\\
     &= \prod_{i=1}^n f(Y_i; \mu_i(R_i, X_i), \Psi)^{A_i} \pi_i(R_i, X_i)^{A_i} (1 - \pi_i(R_i, X_i))^{1-A_i}\nonumber
\end{align*}

We note that when \(F\) admits an absolutely continuous density function
the likelihood function can equivalently be written solely in terms of
the combined outcome \(\widetilde{Y}\) in Equation
(\ref{eq:combinedOutcome}) since
\(f(Y_i; \cdot)^{A_i} = f(\widetilde{Y}_i; \cdot)^{1_{\widetilde{Y}_i \ne \mathcal{E}}}\)
and
\(\pi_i(\cdot)^{A_i} = \pi_i(\cdot)^{1_{\widetilde{Y}_i \ne \mathcal{E}}}\)
almost surely. An important property of this model is that the
likelihood function factorizes into two components as

\begin{align}
 L_n(\mu_i(R_i, X_i), \pi_i(R_i, X_i)) &= L_{1,n}(\mu_i(R_i, X_i), \Psi) L_{2,n}(\pi_i(R_i, X_i))\label{eq:likFactorization}
\end{align}

where
\(L_{1,n}(\mu_i(R_i, X_i), \Psi) = \prod_{i=1}^n f(Y_i; \mu_i(R_i, X_i), \Psi)^{A_i}\)
and
\(L_{2,n}(\pi_i(R_i, X_i)) = \prod_{i=1}^n \pi_i(R_i, X_i)^{A_i} (1 - \pi_i(R_i, X_i))^{1-A_i}\).
This implies that the parameters for the observed outcomes,
\(\mu_i(R_i, X_i)\), can be estimated independently of the parameters
governing the probability of observing the outcome, \(\pi_i(R_i, X_i)\).
This factorization is a consequence of the likelihood construction and
it does neither assume nor require independence between the value of the
outcome and the probability of observing it.

Under the assumption of generalized linear models for
\(\mu_i(R_i, X_i)\) and \(\pi_i(R_i, X_i)\) with additive structures we
may parametrize the mean functions as

\begin{align}
  \E[Y_i \mid A_i = 1, R_i, X_i] &= \mu_0 + \mu_{\delta} R_i + s_\mu(X_i)\label{eq:glm1}\\
  \logit P(A_i = 1 \mid R_i, X_i) &= \beta_0 + \beta_{\delta} R_i + s_\beta(X_i)\label{eq:glm2}
\end{align}

where the treatment effect is quantified by bi-variate contrast
\((\mu_\delta, \beta_\delta)\) and
\(s_\mu, s_\beta\colon\, \mathbb{R}^p \mapsto \mathbb{R}\) are some
models for the baseline covariates. The parameter \(\mu_\delta\) is
interpreted as the expected difference among the observed outcomes and
\(\beta_\delta\) is correspondingly the log odds-ratio of being
observed.

To assess the effect of treatment on the combined outcome we propose a
test statistic based on the likelihood ratio statistic

\begin{align}
  W_n(\mu_0, \mu_\delta, s_\mu, \beta_0, \beta_\delta, s_\beta) &= -2 \log \frac{\sup\limits_{\Psi} L_{1,n}(\mu_0, \mu_\delta, s_\mu, \Psi) L_{2,n}(\beta_0, \beta_\delta, s_\beta)}{\sup\limits_{\mu_0, \mu_\delta, s_\mu, \Psi} L_{1,n}(\mu_0, \mu_\delta, s_\mu, \Psi) \sup\limits_{\beta_0, \beta_\delta, s_\beta} L_{2,n}(\beta_0, \beta_\delta, s_\beta)}\label{eq:lrtgeneral}\\
  &= W_{1,n}(\mu_0, \mu_\delta, s_\mu)W_{2,n}(\beta_0, \beta_\delta, s_\beta)
\end{align}

\begin{align}
  W_n^P(\mu_\delta, \beta_\delta) = \sup\limits_{\mu_0, s_\mu} W_{n,1}(\mu_0, \mu_\delta, s_\mu)\sup\limits_{\beta_0, \beta_s} W_{n.2}(\beta_0, \beta_\delta, s_\beta)  \label{eq:LRTtotal}
\end{align}

use the profile likelihood ratio test (LRT) statistic which can be
written as a function of the treatment effects as and the value of the
LRT statistic under the null-hypothesis of no treatment effect is
therefore \(W_n(0, 0)\). Similar to the factorization of the likelihood
function in Equation (\ref{eq:likFactorization}), the LRT statistic in
Equation (\ref{eq:lrtgeneral}) also decomposes into the sum

\begin{align}
  W_n(\mu_\delta, \beta_\delta) &= W_{1,n}(\mu_\delta) + W_{2,n}(\beta_\delta)\label{eq:LRTsum} 
\end{align}

of two LRT statistics -- one for the continuous part and one for the
discrete part of the combined outcome. Under very general conditions it
follows that \(W_n(\mu_\delta, \beta_\delta)\) is approximately
\(\chi^2\) distributed with two degrees of freedom (Wilks 1938; A. B.
Owen 1988).

In order to actually perform the test we still need to decide on an
stochastic model for the continuous part of the combined outcome,
\(Y_i \mid A_i = 1\). In the following two sections we present both a
parametric approach based on a normal model and a flexible
semi-parametric approach that does not require distributional
assumptions.

\subsection{Parametric approach}\label{parametric-approach}

If we combine the model for the expected value in Equation
(\ref{eq:glm1}) with the assumption of normal distributed outcomes of
the continuous part of the combined outcome we obtain the following
sub-model

\begin{align*}
  Y_i \mid A_i = 1, R_i \sim N\left(\mu_0 + \mu_\delta R_i, \sigma^2\right)
\end{align*}

With the generalized linear models in Equations (\ref{eq:glm1}) and
(\ref{eq:glm2}) the first term is the LRT statistic with a single degree
of freedom in a linear regression model, and the second term is the LRT
statistic with a single degree of freedom in a logistic regression
model. This makes this test very easy to perform in standard statistical
software that outputs the likelihood value for a model fit without the
need of special methods. The calculation of the LRT simply amounts to
estimating two linear regression models and two logistic regression
models -- for each type a model with and a model without the binary
treatment indicator as a covariate. Combining the two likelihood ratios
according to Equation (\ref{eq:LRTsum}) calculates our test statistic,
and the p-value for no treatment effect can be determined through the
\(\chi^2\)-distribution with two degrees of freedom. The critical value
for rejecting the null-hypothesis of no treatment effect at the 5\%
level is equal to \(5.99\) and equal to \(9.21\) at the 1\% level.

A direct consequence of Wilks' theorem (Wilks 1938) is given in the
following proposition. Note that the stated conditions are trivially
satisfied for an RCT with fixed randomization proportions.

\begin{proposition}
Assume that $P(A_i = 1 \mid R_i = r) > 0$ for $r = 0,1$ and let $m_j = \sum_{i=1}^n 1_{R_i = j}$. Then the parametric profile likelihood ratio test statistic $W_n^P(0,0)$ stated in Equation (\ref{eq:LRTtotal}) for the null-hypothesis of no treatment effect on the combined outcome is asymptotically $\chi^2$ distributed with two degrees of freedom for $m_1, m_2 \rightarrow \infty$ and $m_1/m_2 \rightarrow c$ for some finite, positive constant $c$.
\end{proposition}

\subsection{Semi-parametric approach}\label{semi-parametric-approach}

A drawback of the parametric LRT introduced in the previous section is
that it requires deciding on a parametric distribution for the observed
outcomes. In this section we introduce a more flexible approach based on
an empirical LRT. This approach also utilizes the additive decomposition
of the LRT statistic in Equation (\ref{eq:LRTsum}) but substitutes the
term \(W_{1,n}\) with a term that is free of any distributional
assumptions. Combining this with the binomial model for \(A\) leads to a
semi-empirical LRT for the treatment effect.

Let
\(\mathcal{I}_0 = \left\{i = 1,\ldots, n : R_i = 0, A_i = 1\right\}\)
and
\(\mathcal{I}_1 = \left\{i = 1,\ldots, n : R_i = 1, A_i = 1\right\}\) be
the sets of indices of the observed outcomes for the two treatments. The
empirical LRT statistic as a function of the difference in expected
value is given by

\begin{align*}
  W_{1,n}^E(\mu_\delta^\ast) &= 2\sup_{\mu}\left(\sum_{i \in \mathcal{I}_0}\log \left(1 + \lambda_1 (Y_i - \mu)\right) + \sum_{j \in \mathcal{I}_1}\log\left(1 + \lambda_2 (Y_j - \mu - \mu_\delta^\ast)\right)\right)
\end{align*}

where \(\lambda_1\) and \(\lambda_2\) are the solutions to the following
equations

\begin{align*}
  |\mathcal{I}_0|^{-1}\sum_{i \in \mathcal{I}_0} \frac{Y_i - \mu}{1 + \lambda_1 (Y_i - \mu)} = 0, \quad |\mathcal{I}_1|^{-1}\sum_{j \in \mathcal{I}_1}\frac{Y_j - \mu - \mu_\delta^\ast}{1 + \lambda_2 (Y_j - \mu - \beta_\delta^\ast)} = 0
\end{align*}

We refer to the appendix for a derivation of the test statistics.

The semi-parametric LRT statistic for testing the null-hypothesis of no
treatment effect is equal to \(W_n^{SP}(0, 0)\) where

\begin{align}
  W_n^{SP}(\mu_\delta^\ast, \alpha_\delta^\ast) = W_{1,n}^E(\mu_\delta^\ast) + W_{2,n}(\alpha_\delta^\ast) \label{eq:LRT_SP_total}
\end{align}

By the non-parametric Wilk's theorem (A. B. Owen 1988) it follows that
\(W_n^{SP}(0, 0) \sim \chi^2_2\) asymptotically.

\begin{proposition}
Assume that $P(A_i = 1 \mid R_i = r) > 0$ for $r = 0,1$ and let $m_j = \sum_{i=1}^n 1_{R_i = j}$. Then the semi-parametric profile likelihood ratio test statistic $W_n^{SP}(0,0)$ stated in Equation (\ref{eq:LRT_SP_total}) for the null-hypothesis of no treatment effect on the combined outcome is asymptotically $\chi^2$ distributed with two degrees of freedom for $m_1, m_2 \rightarrow \infty$ and $m_1/m_2 \rightarrow c$ for some finite, positive constant $c$.
\end{proposition}

\subsection{Confidence intervals}\label{confidence-intervals}

Confidence intervals for the treatment effects are readily available by
inverting the LRT in Equation (\ref{eq:LRTsum}) utilizing the duality
between hypothesis testing and confidence intervals. Specifically, an
\((1-\alpha)100\%\) confidence region or interval contains all parameter
values that cannot be rejected according to the LRT at level \(\alpha\).
The bi-variate confidence region for the treatment effect is therefore
given by the following point set in \(\mathbb{R}^2\)

\begin{align*}
  \text{CI}^{(\mu_\delta, \beta_\delta)}_{1-\alpha} = \left\{(m, b) :  W_{1,n}(m) +  W_{2,n}(b) \leq \chi_2^2(1-\alpha)\right\}
\end{align*}

and it will asymptotically contain the true difference in means among
the observed outcomes and the true log odds-ratio of being observed
simultaneously with probability \((1-\alpha)100\%\). Similarly,
uni-variate confidence intervals for \(\mu_\delta\) and \(\beta_\delta\)
can be computed by inversion with respect to a \(\chi^2\) distribution
with one degree of freedom, e.g.,

\begin{align*}
  \text{CI}^{\mu_\delta}_{1 - \alpha} = \left\{m : W_{1,n}(m) \leq \chi_1^2(1-\alpha)\right\}, \quad \text{CI}^{\beta_\delta}_{1 - \alpha} = \left\{b : W_{2,n}(b) \leq \chi_1^2(1-\alpha)\right\}
\end{align*}

The exact formula for the confidence interval for the average difference
in the combined outcome (\(\widetilde{Y}\)) depends on the chosen model
for the binary component (\(A\)). However, it can always be calculated
using the Delta-method.

\section{R package}\label{r-package}

To facilitate a straightforward application of our approach we have
implemented both the parametric and semi-parametric likelihood ration
tests in the R package \texttt{TruncComp} which is available at the
first author's GitHub repository (Jensen and Lange 2018). We illustrate
its applicability based on an example data set also available from the
package.

\begin{figure}[htbp]

{\centering \includegraphics{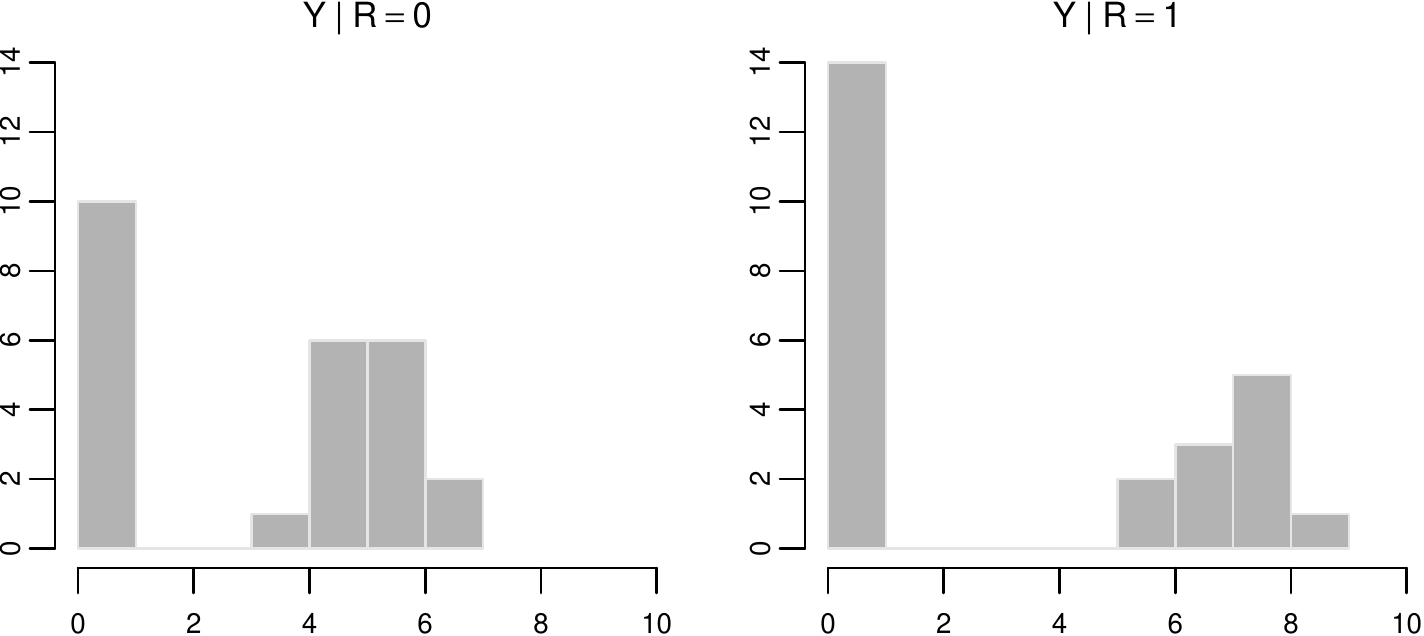} 

}

\caption{\label{fig:dataExampleHistogram}Histograms of the outcome for the example data stratified by treatment group.}\label{fig:unnamed-chunk-2}
\end{figure}

The example data set can be loaded by writing
\texttt{data("TruncCompExample")} after loading the package. The data
set contains two variables, \(Y\) and \(R\), where \(Y\) is the
continuous outcome and \(R\) is the binary treatment indicator. There
are \(25\) observations in each treatment group, and truncated
observations in \(Y\) have been assigned the atom \(\mathcal{E} = 0\).
Figure \ref{fig:dataExampleHistogram} shows histograms of the outcome
for each treatment group. Visually there appears to be a difference
between the two groups both in terms of the frequency of the atom and a
location shift in the continuous part.

The observed difference in means for the combined outcome, \(\Delta\) in
Equation (\ref{eq:Delta}), is 0.018 and both a two-sample t-test and a
Wilcoxon rank sum test show highly insignificant effects of the
treatment with p-values of 0.984 and 0.696 respectively. In order to
analyse the data using proposed method we use the function
\texttt{truncComp} as follows

\begin{Shaded}
\begin{Highlighting}[]
\NormalTok{model <-}\StringTok{ }\KeywordTok{truncComp}\NormalTok{(Y }\OperatorTok{~}\StringTok{ }\NormalTok{R, }\DataTypeTok{atom =} \DecValTok{0}\NormalTok{, }\DataTypeTok{data =}\NormalTok{ TruncCompExample, }\DataTypeTok{method=}\StringTok{"SPLRT"}\NormalTok{)}
\end{Highlighting}
\end{Shaded}

where the formula interface is similar to other regression models
implemented in R. The argument \texttt{atom} identifies the value
assigned to the unobserved outcomes, and \texttt{method} can be
\texttt{SPLRT} or \texttt{LRT} for either the semi-parametric or the
parametric likelihood ratio test, respectively. In this example we have
opted for the semi-parametric version to show how easily this additional
flexibility is included. We obtain the results of the estimation by a
calling the function \texttt{summary} on the estimated model:

\begin{Shaded}
\begin{Highlighting}[]
\KeywordTok{summary}\NormalTok{(model)}
\end{Highlighting}
\end{Shaded}

\begin{verbatim}
## Estimation method: Semi-empirical Likelihood Ratio Test 
## Confidence level = 95%
## 
## Treatment contrasts
##                                          Estimate  CI Lower CI Upper
## Difference in means among the observed: 1.8564296 1.1638863 2.480132
## Odds ratio of being observed:           0.5238095 0.1660407 1.596820
## 
## Joint test statistic: W = 31.09545
## p-value: p = 1.768924e-07
\end{verbatim}

The output from the call to \texttt{summary} displays estimates for the
two treatment contrasts corresponding to \(\mu_\delta\) and
\(e^{\alpha_\delta}\) in Equations (\ref{eq:glm1}) and (\ref{eq:glm2}).
These contrasts quantify the difference in means among the observed
outcomes and the odds ratio of being observed respectively in accordance
with the model specification. Each estimated treatment contrast is
accompanied with a confidence interval, and finally the output displays
the joint likelihood ratio test statistic and the associated \(p\)-value
for the joint null-hypothesis of no treatment effect.

From the output we see that the semi-parametric likelihood ratio
analysis reports an extremely low \(p\)-value for null-hypothesis of no
joint treatment effect. This strongly contradicts the conclusions from
both the previous \(t\)- and Wilcoxon analyses. The confidence intervals
for the two treatment contrasts indicate that the average value for the
observed outcome in the group defined by \(R = 1\) is significantly
higher than the average value for the observed outcome in the group with
\(R = 0\).

The confidence intervals can also be obtained by calling the function
\texttt{confint} on the model object. This command reports both the
marginal confidence intervals for the two treatment contrasts as well as
a their simultaneous confidence region. To obtain the simultaneous
confidence region we write

\begin{figure}[htbp]

{\centering \includegraphics{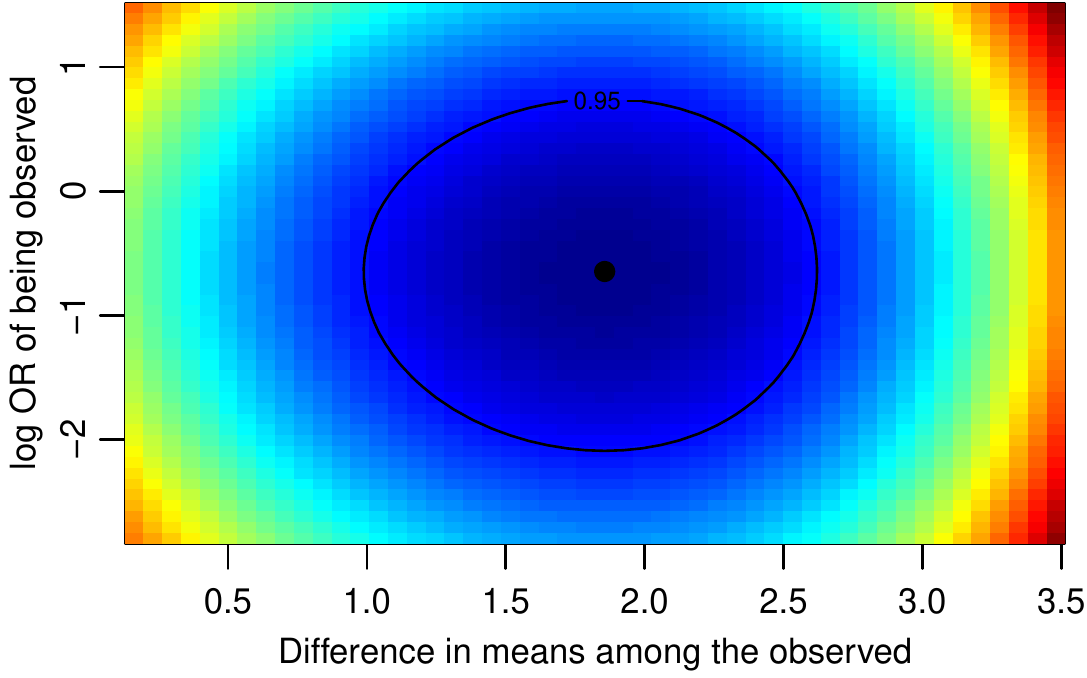} 

}

\caption{\label{fig:simultConfidence}Simultaneous empirical likelihood ratio surface for the two treatment contrasts.}\label{fig:unnamed-chunk-6}
\end{figure}

\begin{Shaded}
\begin{Highlighting}[]
\KeywordTok{confint}\NormalTok{(model, }\DataTypeTok{type=}\StringTok{"simultaneous"}\NormalTok{, }\DataTypeTok{plot=}\OtherTok{TRUE}\NormalTok{, }\DataTypeTok{resolution =} \DecValTok{50}\NormalTok{)}
\end{Highlighting}
\end{Shaded}

where \texttt{resolution} is the number of grid points on which the
surface is evaluated. Figure \ref{fig:simultConfidence} shows a heat-map
of the semi-empirical likelihood surface as well as the \(95\%\)
confidence region. The point \((0,0)\) is far outside of the joint
confidence region which corresponds to the strong rejection of the joint
null hypothesis of nu treatment effect.

\section{Simulation study}\label{simulation-study}

To illustrate the power benefit and small sample properties of our
proposed procedure we consider 4 setups, which we examine by
simulations. In all simulations setups it is assumed that treatment is
randomized 1:1. We will vary both the mean among survivors (denoted
\(\mu_0\) and \(\mu_1\)) and probability of death (\(p_0\) and \(p_1\)).
For the first three simulation setup values among survivors are assumed
to follow a normal distribution with unit variance and the stated means.
We will also consider the effect of heavily right skewed data among the
survivors (simulation setup 4, which uses the square of a t-distribution
with 2 degrees of freedom multiplicatively scaled to having the right
mean. Table 1 below presents the considered simulation setups. It is
noted that in simulation setups 3 and 4 the effect on the survivors and
the effect on mortality are in opposite directions.

\begin{longtable}[]{@{}lrrrrlr@{}}
\caption{Simulation setups. \(\Delta\) denotes the marginal mean when
setting the outcome to zero for the dead.}\tabularnewline
\toprule
& \(\mu_0\) & \(\mu_1\) & \(\pi_0\) & \(\pi_1\) & Distribution &
\(\Delta\)\tabularnewline
\midrule
\endfirsthead
\toprule
& \(\mu_0\) & \(\mu_1\) & \(\pi_0\) & \(\pi_1\) & Distribution &
\(\Delta\)\tabularnewline
\midrule
\endhead
Setup 1 & 3.0 & 4.0 & 0.35 & 0.35 & Normal (sd=1) &
-0.3476400\tabularnewline
Setup 2 & 3.5 & 3.5 & 0.40 & 0.30 & Normal (sd=1) &
0.3481800\tabularnewline
Setup 3 & 3.0 & 4.0 & 0.40 & 0.30 & Normal (sd=1) &
0.0035646\tabularnewline
Setup 4 & 3.0 & 4.0 & 0.40 & 0.30 & t-dist (df=2) squared &
-1.0058211\tabularnewline
\bottomrule
\end{longtable}

To assess power we vary the sample size from 25 to 250 and for each
configuration we conduct 100,000 Monte Carlo replications and compute
power. The resulting power curves are presented in Figure 3. In setup 1
we observe a large power gain compared to the Wilcox test. Here the
Wilcox tests gets ``confused'' by the large number of ties in the atom.
In setup 2 Wilcox test has slightly better power profile. This is to be
expected as our novel test is here disadvantaged by being a
two-degrees-of-freedom test where the Wilcox is only one. It is further
observed that Wilcox as expected as very little power when the effects
on mortality and among survivors are of opposite sign despite the two
distributions being markedly different indicating clear treatment effect
(setups 3 and 4). In contrast our novel methods has excellent power. In
all settings with a normally distributed outcome among survivors the
parametric and semi-parametric approaches are similar, but for the heavy
tail setup (no. 4) the semi-parametric approach is clearly superior.

\begin{figure}[htbp]

{\centering \includegraphics{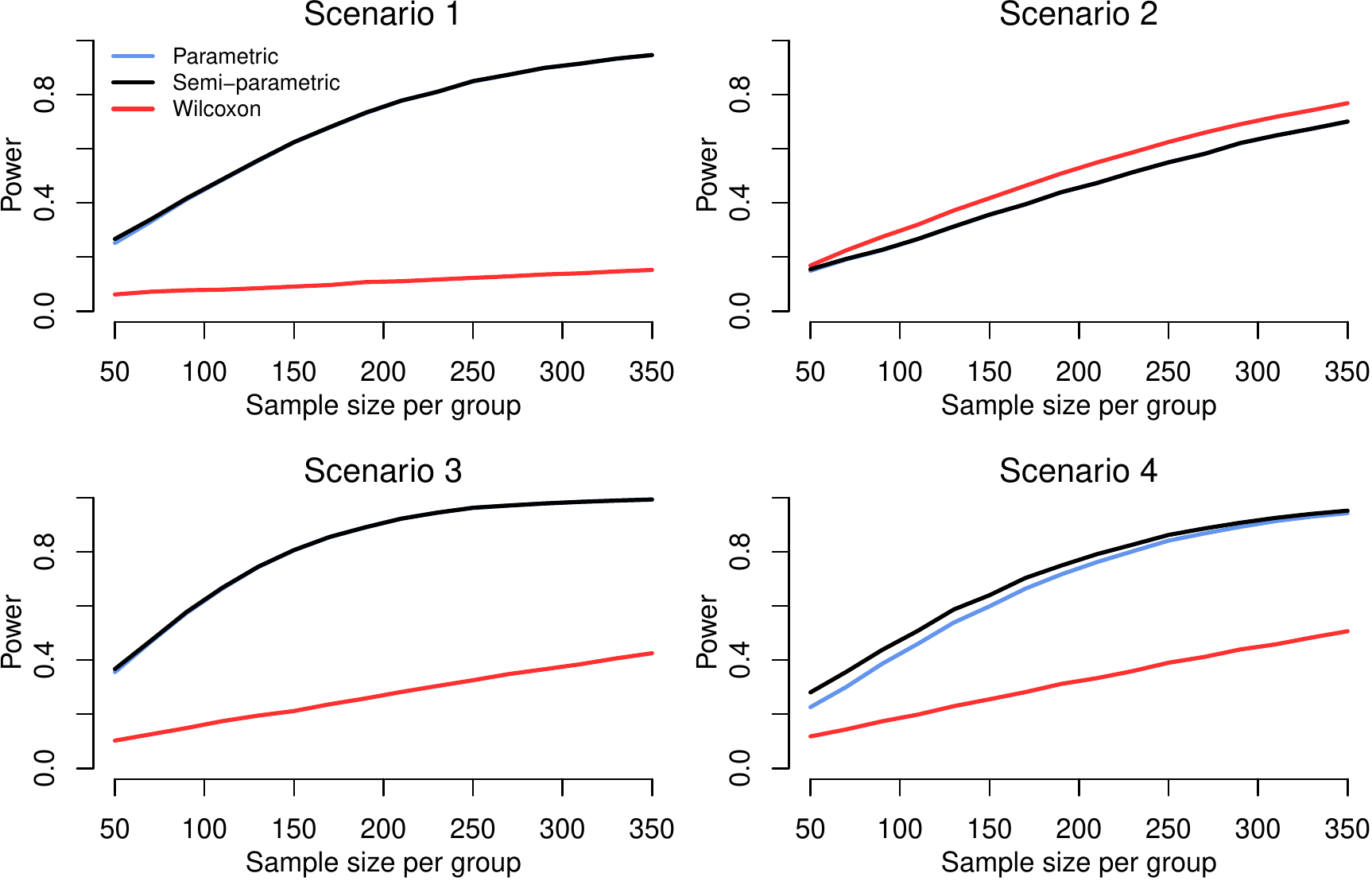} 

}

\caption{\label{fig:powerCurves}Simulated power as a function of sample size for each of the four scenarios in Table 1.}\label{fig:unnamed-chunk-8}
\end{figure}

\section{Discussion}\label{discussion}

In this paper we introduce a novel statistical test to assess treatment
effect on continuous outcomes where one value has special meaning (e.g.,
all diseased are assigned lowest possible value). The procedure in
potentially much more power-full than the current best-practice which is
to use Wilcox-type tests. The proposed method includes both a fully
parametric approach and a semi-parametric where one makes no assumptions
on the functional form of the continuous part of the distribution. In
all settings the new method not only provide an effect measure but also
effect parameters with associated confidence intervals. The test is
implemented in an R package available on GitHub.

It is noted that unlike the Wilcox test out proposed method can easily
be extended to include covariates (A. Owen 1991). It is therefor not
only useful in an RCT setting but also to observed data.

\section*{References}\label{references}
\addcontentsline{toc}{section}{References}

\hypertarget{refs}{}
\hypertarget{ref-TruncCompGitHub}{}
Jensen, Andreas Kryger, and Theis Lange. 2018. ``The TruncComp R package
for Two-Sample Comparison of Truncated Continuous Outcomes Using
Parametric and Semi-Empirical Likelihood.''
\url{https://github.com/aejensen/TruncComp}.

\hypertarget{ref-owen1991empirical}{}
Owen, Art. 1991. ``Empirical Likelihood for Linear Models.'' \emph{The
Annals of Statistics}, 1725--47.

\hypertarget{ref-owen1988empirical}{}
Owen, Art B. 1988. ``Empirical Likelihood Ratio Confidence Intervals for
a Single Functional.'' \emph{Biometrika} 75 (2): 237--49.

\hypertarget{ref-wilks1938large}{}
Wilks, Samuel S. 1938. ``The Large-Sample Distribution of the Likelihood
Ratio for Testing Composite Hypotheses.'' \emph{The Annals of
Mathematical Statistics} 9 (1): 60--62.

\appendix

\newpage

\section*{Appendix}\label{appendix}
\addcontentsline{toc}{section}{Appendix}

\subsection*{Derivation of semi-parametric test
quantity}\label{derivation-of-semi-parametric-test-quantity}
\addcontentsline{toc}{subsection}{Derivation of semi-parametric test
quantity}

The empirical likelihood ratio function that compares the empirical
maximum likelihood under a constraint set \(\mathcal{C}\) to an
unconstrained maximum likelihood is given by

\begin{align}
R_{1,n}^\text{E}(\mathcal{C}) &=  \frac{\sup\limits_{\left\{p_i\right\}}\left(\prod_{i \in \mathcal{I}_0} p_i \mid \mathcal{C}\right) \sup\limits_{\left\{q_j\right\}}\left(\prod_{j \in \mathcal{I}_1} q_j \mid \mathcal{C} \right)}{\sup\limits_{F \in \mathcal{F}}\prod_{i \in \mathcal{I}_0} F(\left\{Y_i\right\}) \sup\limits_{G \in \mathcal{F}}\prod_{j \in \mathcal{I}_1} G(\left\{Y_j\right\})}
\end{align}

where \(\mathcal{F}\) is the family of cadlag functions. This empirical
likelihood ratio is simply a comparison between the non-parametric
unconstrained maximum likelihood values and a a null-model where the
maximum likelihoods in the two treatment groups are constrained
according to a constraint set \(\mathcal{C}\). The sets of weights,
\(\left\{p_i\right\}\) and \(\left\{q_j\right\}\), form a constrained
multinomial distribution over the observations.

It is well-known that the solutions to the two unconstrained
maximizations in the denominator are given by the empirical distribution
functions that put equal weight on each observation. From here it
follows that the likelihood ratio function can be written as

\begin{align}
R_{1,n}^\text{E}(\mathcal{C}) &= \sup\limits_{\left\{p_i\right\}, \left\{q_j\right\}} \left(\prod_{i \in \mathcal{I}_0} |\mathcal{I}_0|p_i  \prod_{j \in \mathcal{I}_1} |\mathcal{I}_1|q_j \mid \mathcal{C}\right)\label{eq:empiricalLik1}
\end{align}

where \(|\cdot|\) denotes the cardinality of the index set.

The constraint set is chosen so that \(\left\{p_i\right\}\) and
\(\left\{q_j\right\}\) are bona fide multinomial distributions, and so
that the expected value of the observations with indices in
\(\mathcal{I}_0\) have expectation \(\beta\) and the difference between
the expectations comparing the observations with indices in
\(\mathcal{I}_1\) to those with indices in \(\mathcal{I}_0\) is given by
\(\beta_\delta\) similar to the structure of the linear model in
Equation (\ref{eq:glm2}). This yields the following set of constraints:

\begin{alignat}{3}
 p_i &\geq 0,  &\quad  \sum_{i \in \mathcal{I}_0}p_i &= 1,  &\quad &\sum_{i \in \mathcal{I}_0}p_i (Y_i - \beta) = 0\\   
 q_j &\geq 0,  &\quad  \sum_{j \in \mathcal{I}_1}q_j &= 1,  &\quad &\sum_{j \in \mathcal{I}_1}q_j (Y_j - \beta - \beta_\delta) = 0
\end{alignat}

The find the values of \(\left\{p_i\right\}\) and \(\left\{q_j\right\}\)
in Equation (\ref{eq:empiricalLik1}) that satisfies the constraint set
we solve the constrained optimization problem through the following
objective function

\begin{align}
  O(\left\{p_i\right\}, \left\{q_j\right\}) &= \sum_{i \in \mathcal{I}_0}\log(|\mathcal{I}_0|p_i) + \sum_{j \in \mathcal{I}_1}\log(|\mathcal{I}_1|q_j) + \gamma_1\left(\sum_{i \in \mathcal{I}_0} p_i - 1\right) + \gamma_2\left(\sum_{j \in \mathcal{I}_1}q_j - 1\right)\\
    &- |\mathcal{I}_0|\lambda_1\sum_{i \in \mathcal{I}_0} p_i(Y_i - \beta) - |\mathcal{I}_1|\lambda_2\sum_{j \in \mathcal{I}_1} q_j(Y_j - \beta - \beta_\delta)\nonumber
\end{align}

where \(\gamma_1\), \(\gamma_2\), \(\lambda_1\) and \(\lambda_2\) are
Lagrange multipliers.

Calculating the partial derivatives of
\(O(\left\{p_i\right\}, \left\{q_j\right\})\) with respect to \(p_i\)
and \(q_i\) and setting them equal to zero we have that

\begin{align}
  0 &= \frac{\partial}{\partial p_i}O(\left\{p_i\right\}, \left\{q_j\right\}) = \frac{1}{p_i} - |\mathcal{I}_0|\lambda_1 (Y_i - \beta) + \gamma_1\label{eq:lambda1}\\
  0 &= \frac{\partial}{\partial q_j}O(\left\{p_i\right\}, \left\{q_j\right\}) = \frac{1}{q_j} - |\mathcal{I}_1|\lambda_2 (Y_j - \beta - \beta_\delta) + \gamma_2\label{eq:lambda2}
\end{align}

and by applying the constrains it follows that

\begin{align}
  0 &= \sum_{i \in \mathcal{I}_0} p_i \frac{\partial O(\left\{p_i\right\}, \left\{q_j\right\})}{\partial p_i}\\
   &= \sum_{i \in \mathcal{I}_0} p_i\left(\frac{1}{p_i} - |\mathcal{I}_0|\lambda_1 (Y_i - \mu) + \gamma_1\right)\nonumber\\
   &= \sum_{i \in \mathcal{I}_0} 1  - |\mathcal{I}_0| \lambda_1\sum_{i \in \mathcal{I}_0}p_i (Y_i - \beta) + \gamma_1 \sum_{i \in \mathcal{I}_0} p_i\nonumber\\
   &= |\mathcal{I}_0| + \gamma_1\nonumber
\end{align}

Therefore \(\gamma_1 = -|\mathcal{I}_0|\) and
\(\gamma_2 = -|\mathcal{I}_1|\) by similar calculation for \(q_j\). From
Equations (\ref{eq:lambda1}) and (\ref{eq:lambda2}) and the previous
results we obtain the following solutions

\begin{align}
  p_i = \frac{1}{|\mathcal{I}_0|(1 + \lambda_1 (Y_i - \beta))}, \quad q_j = \frac{1}{|\mathcal{I}_1|(1 + \lambda_2 (Y_j - \beta - \beta_\delta))}
\end{align}

and by inserting these expressions into the expression for the empirical
likelihood ratio we obtain

\begin{align}
\log R_{1,n}^E(\beta, \beta_\delta) &= \sum_{i \in \mathcal{I}_0} \log(|\mathcal{I}_0| p_i) + \sum_{j \in \mathcal{I}_1}\log(|\mathcal{I}_1| q_j)\\
                    &= \sum_{i \in \mathcal{I}_0}\log \frac{1}{1 + \lambda_1 (Y_i - \beta)} + \sum_{j \in \mathcal{I}_1}\log \frac{1}{1 + \lambda_2 (Y_j - \beta - \beta_\delta)}\nonumber\\
                    &= -\sum_{i \in \mathcal{I}_0}\log \left(1 + \lambda_1 (Y_i - \mu)\right) - \sum_{j \in \mathcal{I}_1}\log\left(1 + \lambda_2 (Y_j - \beta - \beta_\delta)\right)\nonumber
\end{align}

The values of \(\lambda_1\) and \(\lambda_2\) can then be found by
combining Equations (\ref{eq:lambda1}) and (\ref{eq:lambda2}) with the
constraints leading to the following set of equations

\begin{align}
  |\mathcal{I}_0|^{-1}\sum_{i \in \mathcal{I}_0} \frac{Y_i - \beta}{1 + \lambda_1 (Y_i - \beta)} = 0, \quad |\mathcal{I}_1|^{-1}\sum_{j \in \mathcal{I}_1}\frac{Y_j - \beta - \beta_\delta}{1 + \lambda_2 (Y_j - \beta - \beta_\delta)} = 0
\end{align}

that in practice can be solved for \(\lambda_1\) and \(\lambda_2\) using
a numerical root finding method.

The empirical LRT statistic is therefore

\begin{align}
  W_{1,n}^E(\beta, \beta_\delta) &= -2\log R_{1,n}^E(\beta, \beta_\delta)\\
                 &= 2\left(\sum_{i \in \mathcal{I}_0}\log \left(1 + \lambda_1 (Y_i - \beta)\right) + \sum_{j \in \mathcal{I}_1}\log\left(1 + \lambda_2 (Y_j - \beta - \beta_\delta)\right)\right)\nonumber
\end{align}

where
\(W_{1,n}^E(\beta_\delta^\ast) = \sup\limits_{\beta} W_{1,n}^E(\beta, \beta_\delta^\ast)\)
the profile version testing the difference between treatments.

\end{document}